# Experimentally Validated Bond Graph Model of a Brazed-Plate Heat Exchanger (BPHE).

M.Turki*, M. Kebdani*, G. Dauphin-Tanguy*, A. Dazin**, P. Dupont***.
* Ecole Centrale de Lille/ CRIStAL UMR CNRS 9189, CS 20048, 59651 Villeneuve d'Ascq. France.
marwa.turki@ec-lille.fr; mohamed.kebdani@ec-lille.fr; genevieve.dauphin-tanguy@ec-lille.fr.
** Arts et Métiers Paris Tech/ LML UMR CNRS 8107, Boulevard Louis XIV, 59000 Lille. France.
antoine.dazin@ensam.eu.
*** Ecole Centrale de Lille/LML UMR CNRS 8107, CS 20048, 59651 Villeneuve d'Ascq. France.
patrick.dupont@ec-lille.fr.

**Keyword:**
Fluid loop, heat exchanger, bond graph, dynamic, single phase flow.

**Abstract -** The paper deals with the Bond Graph (BG) modeling and the model validation of a brazed-plate heat exchanger. This device is an important part of a mechanically pumped cooling loop. A thermo hydraulic BG model is developed and compared with experimental data. Optimization is performed to determine the best value of the convection heat exchange coefficients to be fixed in the model.

## I. INTRODUCTION

Power components dissipate heat flows which represent a significant source of heat. If it is not processed properly, this results in problems that lead to a complete dysfunction of the component. Several solutions have been proposed to prevent the damage of embedded devices and ensure their proper functioning.

Based on these solutions, the choice of the enterprises has been oriented towards mechanically pumped cooling loops that are undoubtedly the most efficient in terms of heat transfer (Kebdani et al [1]). Indeed, in addition to their architectural flexibility, they guarantee appropriate cooling in suitable operational conditions. It is a very important property especially when we know that the continuous cycling of the temperature has a negative impact on the electronic components' life.

The **figure 1** shows up the considered cooling loop. It is composed of a pump, a pressure regulator, an evaporator, a condenser and pipes.

Various heat exchanger technologies exist; the choice depends on the nature of intended use. For example, for space activities, one can use a radiator; for land-based activities (automotive, rail...) an air exchanger with air cross-flow may be adequate.

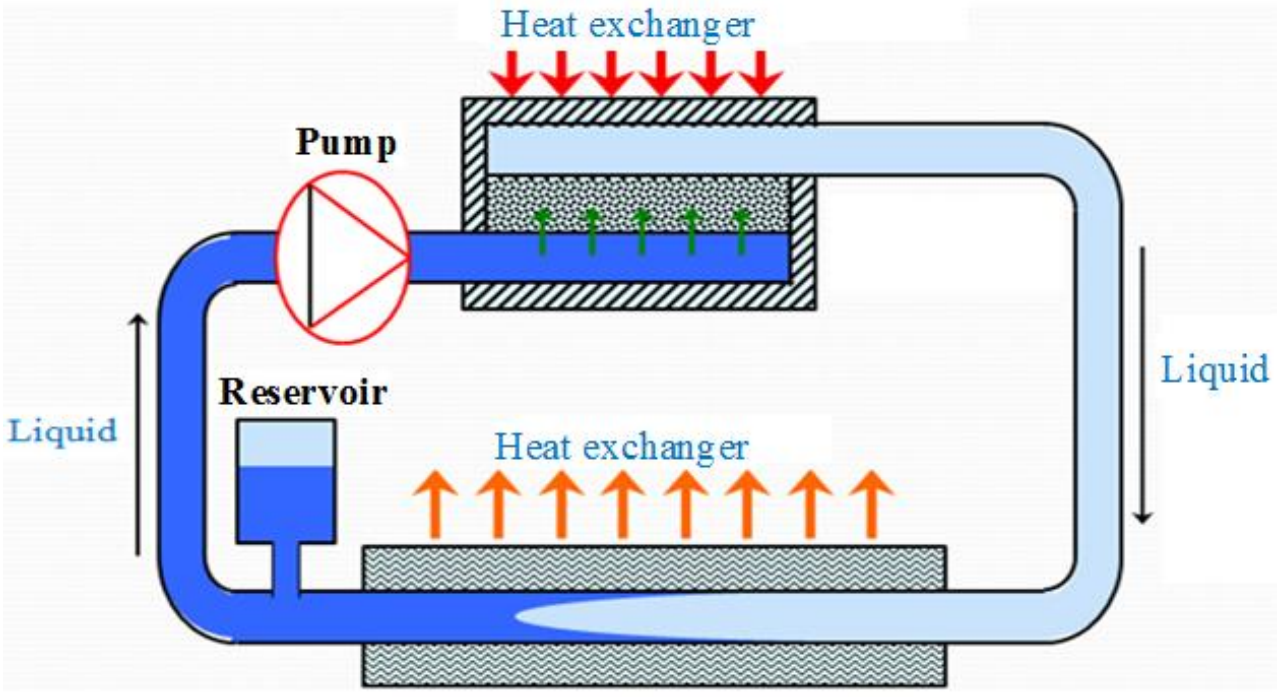

***Fig.1**: Design of the cooling loop*

A single-phase fluid loop is a heat transfer circuit. It is generally a closed loop, wherein a fluid is initially in the liquid state. The fluid is heated to store thermal flux from a heat source, and is then transferred to a heat sink where the hot fluid is cooled and even sub cooled releasing heat to a cold source. Such fluidic loops have high cooling efficiency.

For our purpose, we have chosen to work with the condenser SWEP of reference: "B5Tx6" (***Fig.2***). See Condensers technical specification [2].

The BPHEs have been used for the first time in the 1930s. They were mostly integrated as mono-phasic (liquid-to-liquid) exchangers in the food industry [3] and in the heat pumps [4] thanks to their multiple benefits listed below:

- Optimized effective exchange surface.
- Rational distribution of the flow in the channels.
- Good adaptability.
- High compacity.

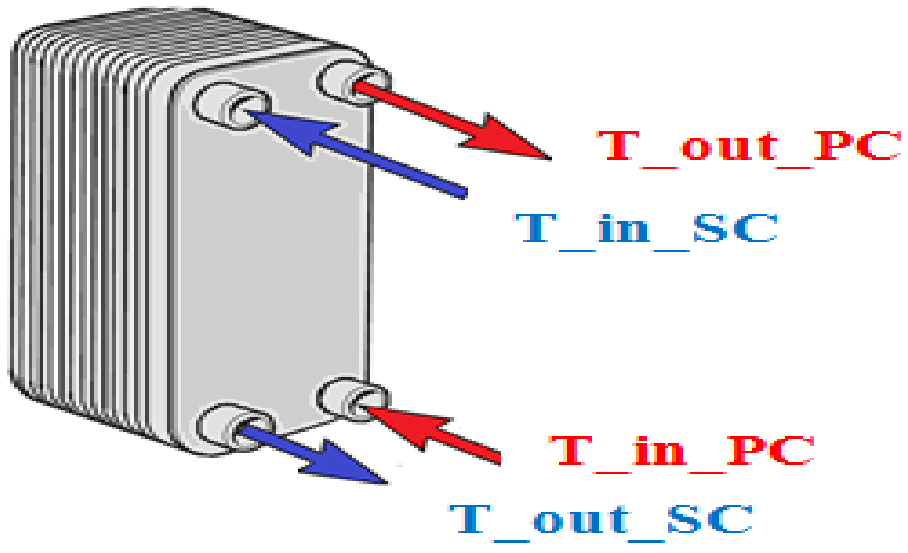

***Fig.2**: Design of the condenser*

Also, the corrugated structure of the thermal plates and chevrons promote turbulent flow (Focke et al. [5]). This leads to high heat transfer efficiency and consequently better performance of the whole installation.

This paper is organized into five sections; the first section is a presentation of our work's overall context and focuses on what has already been done in the BG condenser modeling. The second part deals with the BG model proposed by introducing various assumptions taken into account and the equations put in. The third section is dedicated to the experimental set-up and the various tests performed. The fourth section illustrates an optimization of the heat transfer coefficient. The last part concerns the validation of the proposed model with experimental results

## State of the art

The evaluation of such exchanger's performance is not easy as it needs the development of specific experimental methods, especially for prediction of the exchange coefficient and pressure drops. In this context, various authors conducted studies of instrumentation and visualization:

 Among the works related to this type of heat exchangers, there have been attempts based on the observation of two-phase water-air flows (Vlasogiannis et al. [6] and Volker & Kabelac [7]) for a plate heat exchanger formed of a single channel constituted with a transparent plate. Consequently a flow pattern map is performed.

 More recently Freund & Kabelac [8] have developed an experimental technique based on infrared visualization, to characterize the spatial distribution of the convective heat exchange coefficient for a single-phase flow (water).

 Xiaoyang et al [9] declare that it is possible to estimate theoretically the performance of a BPHE operating in single phase with water as refrigerant fluid. The final results of their work show that it would be appropriate to utilize correlations of pressure drop and heat transfer that take into account corrugation chevron angles.

 Performance of plate heat exchangers under single phase operations are extensively inspected since the 1980s (Cooper and Usher, 1983 [10]; Raju and Bansal, 1983 [11]; Focke et al., 1985 [12]; Shah and Focke, 1988 [13]; Bansal and Muller-Steinhagen, 1993 [14]).
According to these authors, these theoretical works provide a reliable basis that ensures an efficient sizing of the exchanger.

 B. Ould Bouamama (1997) [15] proposes a model, based on BG methodology, of a simple tubular condenser. The mathematical formulation of the problem is clearly described and used to generate a BG intending to predict the dynamic behavior of the condenser operating under two-phase conditions.

In our knowledge, there are no other attempts in the published literature which aim to model BPHEs by referring to the BG approach.

The main purpose of this paper is to propose a validated thermo hydraulic BG model of a chevron type BPHE, operating under single-phase conditions, where the dynamic regime is considered. Simulation results are validated using the test rig developed by the French enterprise Atmostat.

Moreover, according to the research mentioned above, the correlations of the heat exchange are strongly related to experimental conditions in which they were developed. This means that for our case it would be legitimate to pick up the most suitable correlation and, thereafter, perform an optimization study in order to minimize the difference between experimental results and those provided by the model.

Given the nature of the device which involves several physical domains (thermal, hydraulic, electric ...), the bond graph methodology appeared as an appropriate modeling tool.

## II. Bond graph model of the condenser

The thermal fluid processes involve two phenomena: convection, describing the heat transfer between the fluid in motion and walls, and heat conduction, representing the thermal power transmitted due to a temperature difference.
Furthermore, the friction effect in the fluid in movement contributes to the heating of the fluid, especially in case of complicated geometries. Then, it would be wise to analyze the thermo hydraulic coupling for a better modeling of the brazed-plate heat exchanger. This is the purpose of the present section.

### 1) Generalized variables used.

- ⇨ ***Hydraulic part:*** Pressure "$P$" and mass flow rate "$\dot{m}$" for which the product is not a power. The developed model is a "pseudo bond graph".

- ⇨ ***Thermal part:*** for a fluid in motion the entropy balance is not conservative, which justifies the necessity to choose the heat balance instead.

### 2) Coupled Bond Graph of the condenser.

The condenser is mainly composed of a (hot) primary circuit (PC) and a (cold) secondary circuit (SC) separated by brazed- plates as shown **Fig.3**.

***Assumptions:***

1) The internal geometry of the brazed-plate condenser is too complicated to be precisely modeled. Thus we simply consider that such exchanger has a simple rectangular section $(e*e)$ without any corrugations.

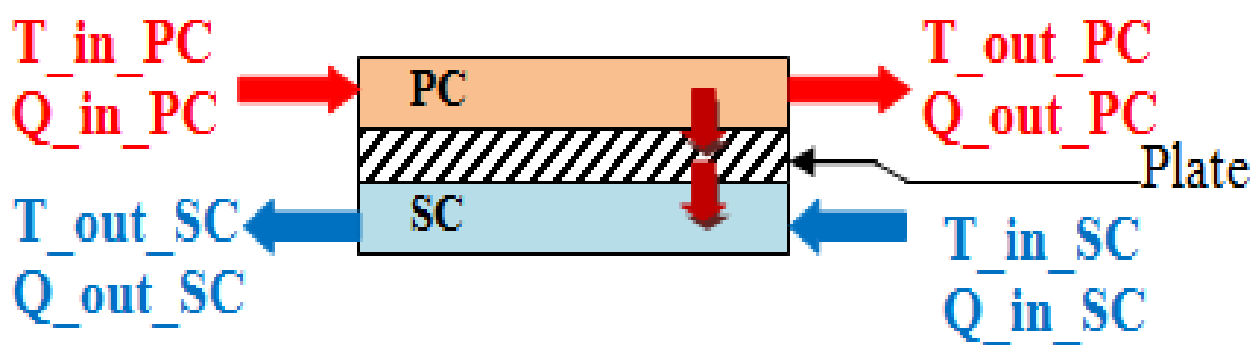

***Fig.3**: longitudinal cup of the condenser*

2) The upstream pressure variation is known and modeled as **MSe**: effort source **Fig.4**.
3) The device is not correctly isolated from the ambient, which is modeled as a MSe.
4) The flow is supposed to be constantly monophasic.
5) Working fluid and secondary fluid are pure water.
6) Each part of the condenser (PC and SC) is modeled as a unique volume where the phenomena are supposed to be homogeneous.

The figure 4 shows up the bond graph model of the condenser.

## *I. RC-elements*

The hydraulic power transfer is represented **Fig.4** by blue half arrows, while the thermal part appears in orange half arrows. The coupling element **RC** represents both the pressure losses generated by walls friction (R part) and the thermal energy storage phenomenon in the exchanger (thermal capacitance effect). The equations describing these phenomena are:

**Hydraulic part**:

For this part, two kinds of pressure losses are taken into account in the dynamic model:

- *Linear pressure drop:*

$$\Delta P_1 = \frac{1}{2} * \rho * v^2 * f * \frac{l}{e} \quad (1)$$

where the Darcy coefficient « f » depends on the flow regime as shown in table.1.

| | |
|---|---|
| Re <2100 **Laminar regime** | $f = \frac{64}{Re}$ |
| 2100<Re <3000 **Transient regime** | $f = \frac{0.316}{Re^{0.25}}$ |
| Re >3000 **Turbulent regime** | $f = \frac{1}{(1.81 lgRe - 1.5)^2}$ |

***Table 1**: Darcy coefficient as a function of the flow regime.*

The Reynolds number $\mathrm{Re} = \frac{\rho * v * e}{\mu}$ is a function of the flow velocity $v = \frac{\dot{m}}{\rho * e * e}$.

- *Singular pressure losses due to two elbows:*

$$\Delta P_2 = 2 * [\frac{1}{2} * \rho * v^2 * \zeta] \quad (2)$$

For an elbow with sharp angle "90°" one can take ζ= 1.3.

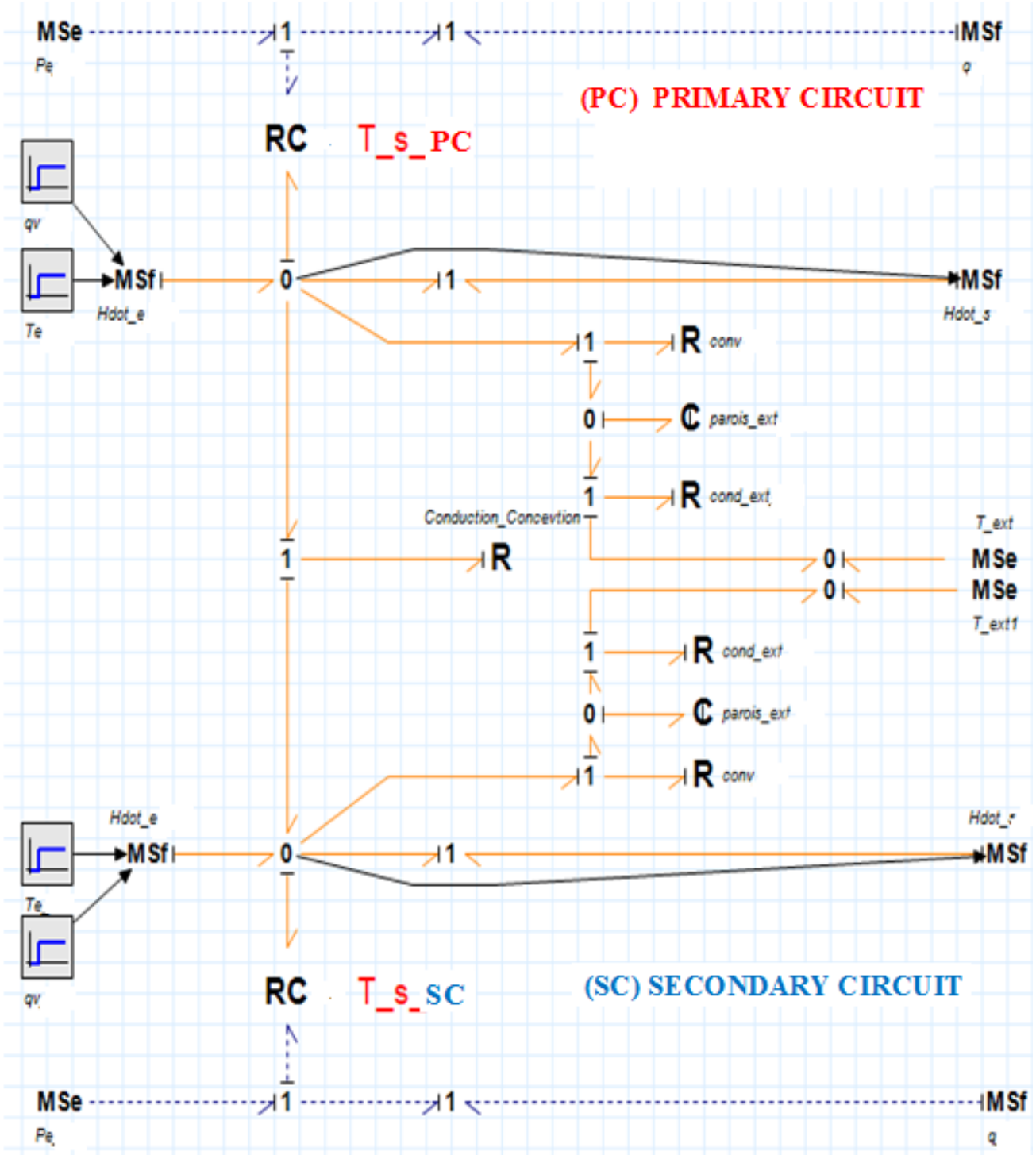

***Fig.4** Pseudo Bond Graph model of the condenser.*

**Thermal part**:

- The heat flow due to the total pressure losses lead to

$$\dot{Q} = (\Delta P_1 + \Delta P_2) * \frac{\dot{m}}{\rho} \quad (3)$$

- Temperature in each part of the condenser is given by

$$T = \frac{H_{cond}}{m_0 * Cp} \quad (4)$$

*with*

$$H_{cond} = \int ((\dot{H}_e - \dot{H}_s) + \dot{Q}) dt + H_0 \quad (5)$$

with initial enthalpy : $H_0 = m_0 * Cp * T_0$

and initial mass $m_0 = \rho_{liq} * V_{cond}$

## *II. R- elements*

The two elements R: conv deal with the heat exchange by convection between each working fluid and the brazed-plate that separate the two fluids.

The corresponding heat flow is:

$$\dot{Q}_{conv} = h \times S \times \Delta T \quad (6)$$

The two elements R: cond_ext represent the heat transfer by conduction within the wall of the condenser to the ambient.

The corresponding heat flow is:

$$\dot{Q}_{cond} = K_{equiv} \times S \times \Delta T \quad (7)$$

The problem is now to experimentally determine the values of the convection coefficients $h_{PC}$ and $h_{SC}$.

$$K_{equiv} = \frac{1}{\frac{1}{h_{cond}} + \frac{1}{h_{CP}} + \frac{1}{h_{CS}}} \quad (8)$$

## III. Experimental set-up

The set-up presented in **Fig.5** has been designed by the French company Atmostat. It is composed of all the devices of the cooling loop.
The test bench was equipped with pressure, temperature, and flow rate sensors. All the experiments have been realized in single-phase state of the fluids.

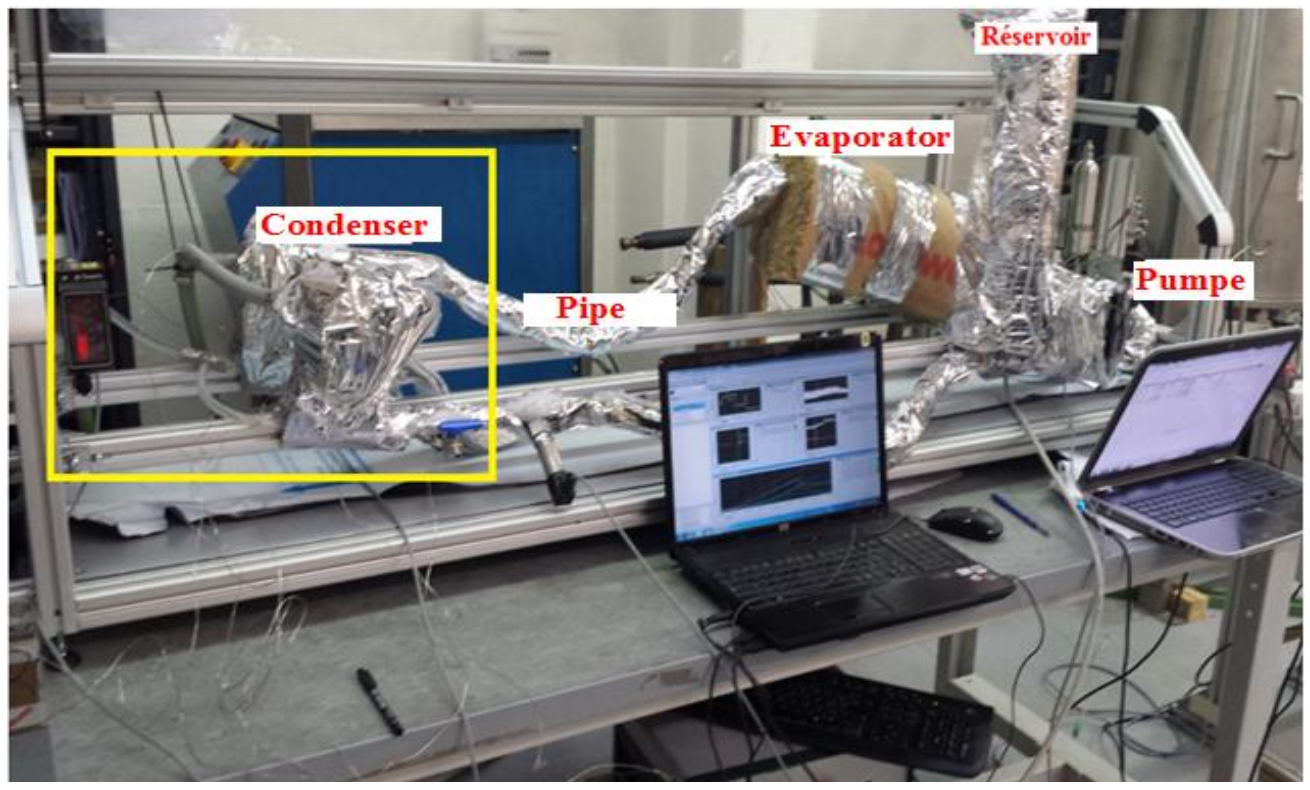

***Fig.5:** Design of the real loop*

### *a) Hydraulic analysis.*

In view of validating the thermo-hydraulic model of the condenser, we consider here the following monophasic test:
$at: \ t = 0s; \quad Q_{v_{CP}} = 5.09 \ cm^3/s.$
$at: \ t = 640s; \ \ Q_{v_{CS}} = 7.44 \ cm^3/s.$

***<u>Analysis of the Figure 6:</u>***

Simulation results are compared with experimental results for downstream liquid pressure in **Fig.6 (1).** These results show good agreement between the model and the real hydraulic behavior, with a discrepancy of $\frac{\Delta P\exp - \Delta P\text{mod}}{\Delta P\exp} = 3\%$ **Fig.6 (2).** The minor discrepancies between the results could be explained by the approximate modeling of the real condenser geometry.

### *b) Thermal analysis.*

***<u>Effect of the total pressure losses:</u>***

It appears that for the actual monophasic test, the total pressure losses (about ***900 Pa)*** calculated by the model $(\Delta P_1 + \Delta P_2)$ are leading to very low heating $\dot{Q}$ (about ***0.0045 W*** ) that could be totally ignored**.**

In fact, it can be argued that as long as the flow is purely monophasic, the heating $\dot{Q}$ due to the friction may be omitted as shown in **Fig.7** where wall temperatures are exactly the same with and without thermo hydraulic coupling in PC, **Fig.7 (a)** and in SC, **Fig.7 (b).**

However, a biphasic study of the condenser will need to take into account the thermal-hydraulic coupling.

***<u>Comparison between fluid and wall temperature:</u>***

At the steady-state the difference between the core working fluid temperature (model), **Fig.8 (a)** green curve, and wall temperature (model), **Fig.8 (a)** red curve, is about **1°C** while the difference between the secondary fluid (model), **Fig.8 (b)** green curve, and the wall temperature (model), **Fig.8 (b)** blue curve, is about **1.3°C.**
The temperature sensors on the set-up are fixed on the tube wall, which makes it impossible to have any information on the temperature inside the fluid. Thus in the following, the comparison between model and experiment results deal with the wall temperatures

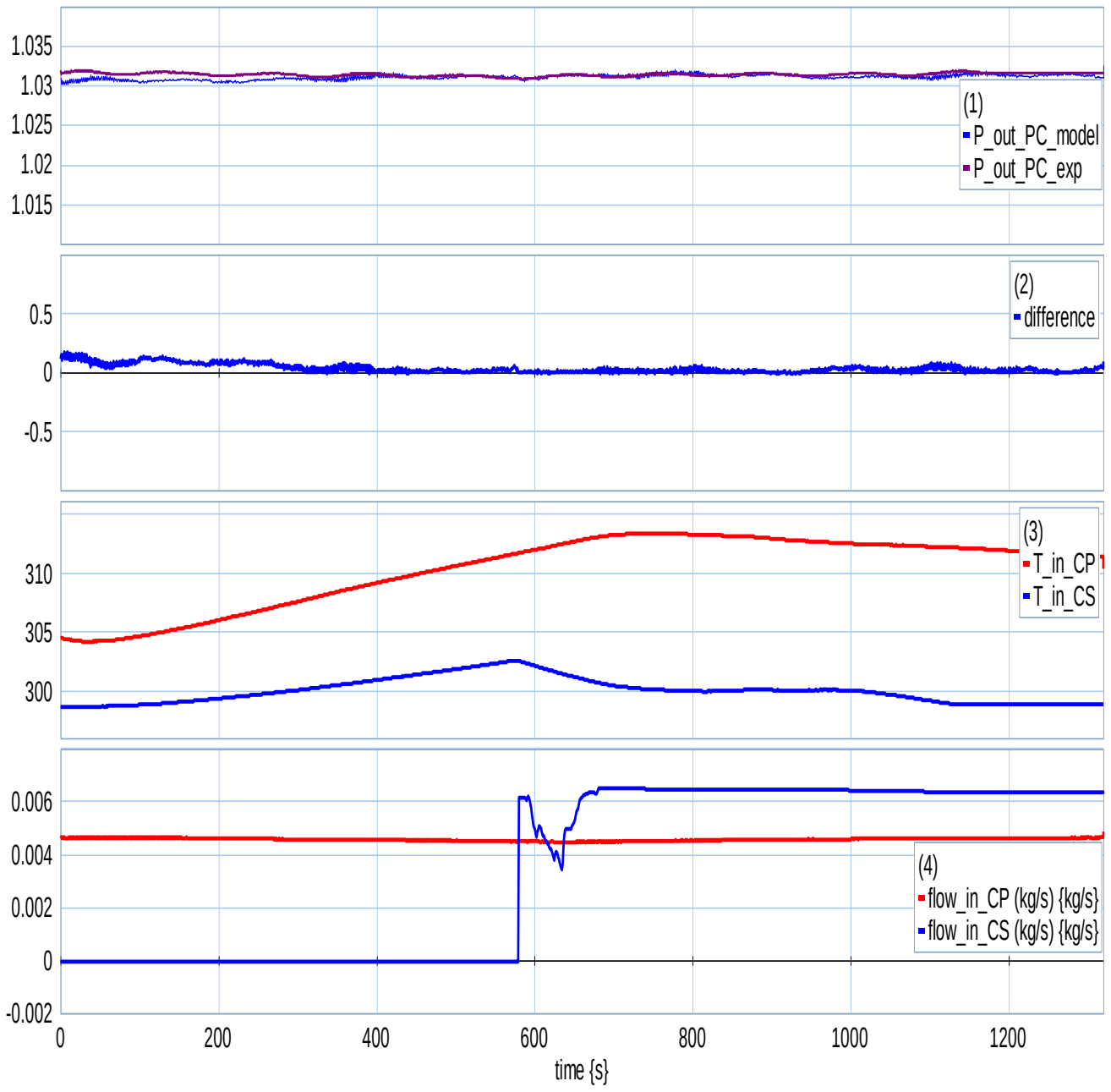

***Fig.6** (a) Time evolution of the outlet pressure of the condenser. (b) Discrepancy between experimental pressure drop and modeled one.*

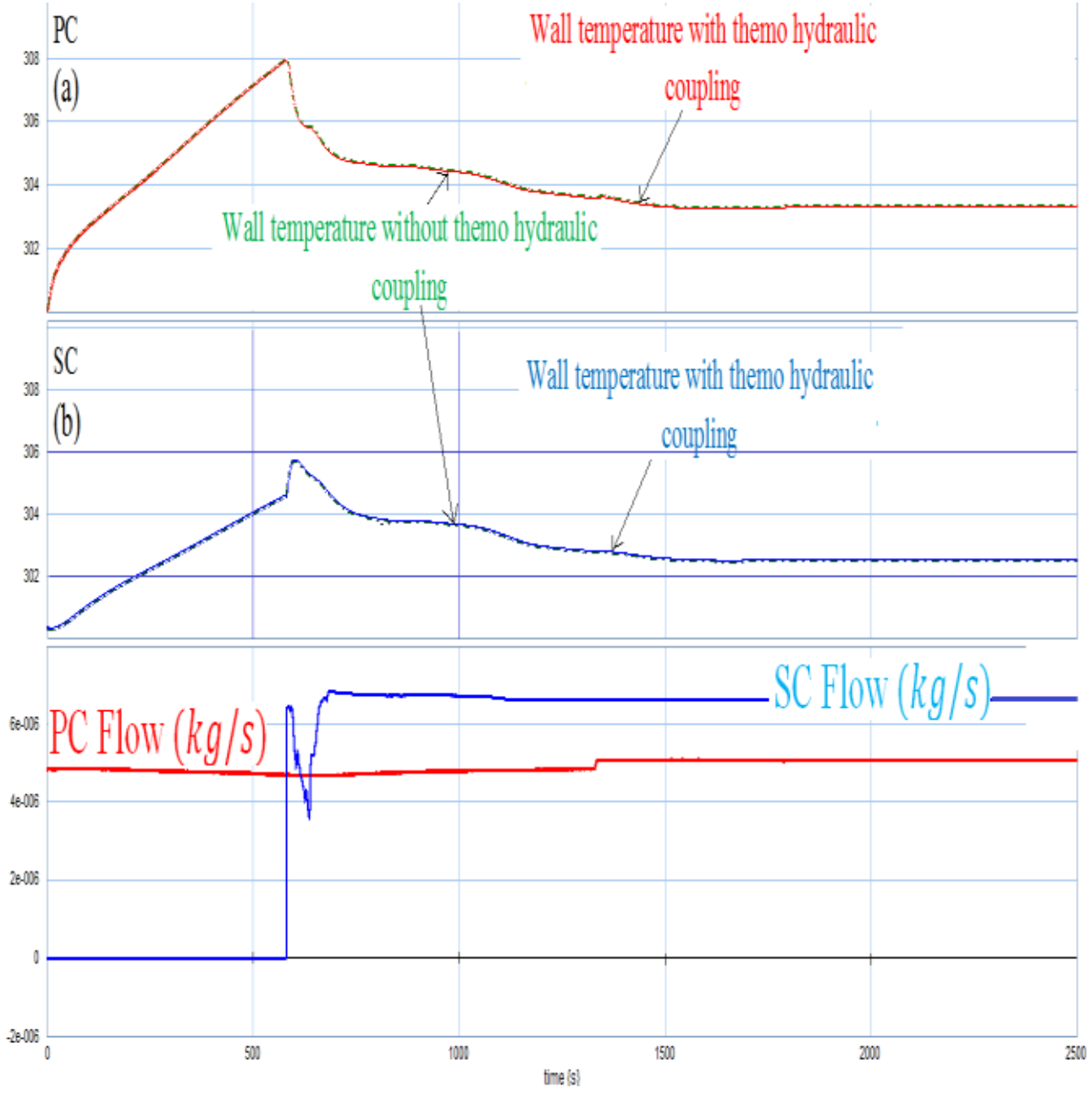

***Fig.7**: Temperatures; in the PC (Graph a), in the SC (Graph b), time evolution of input flow in PC and SC (Graph c).*

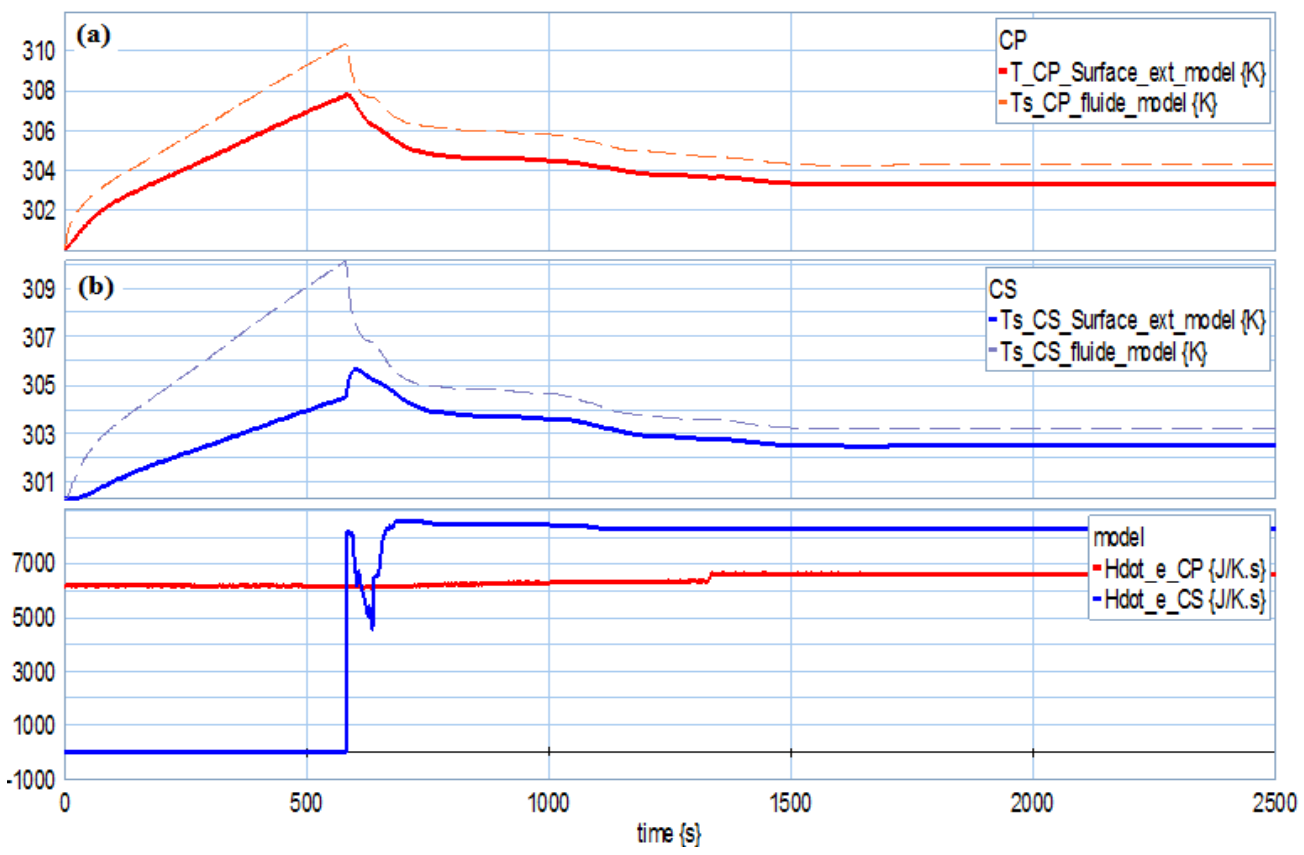

***Fig.8:*** *Time evolution of fluid temperature in PC (a), in SC (b), and flow evolution in both PC and SC.*

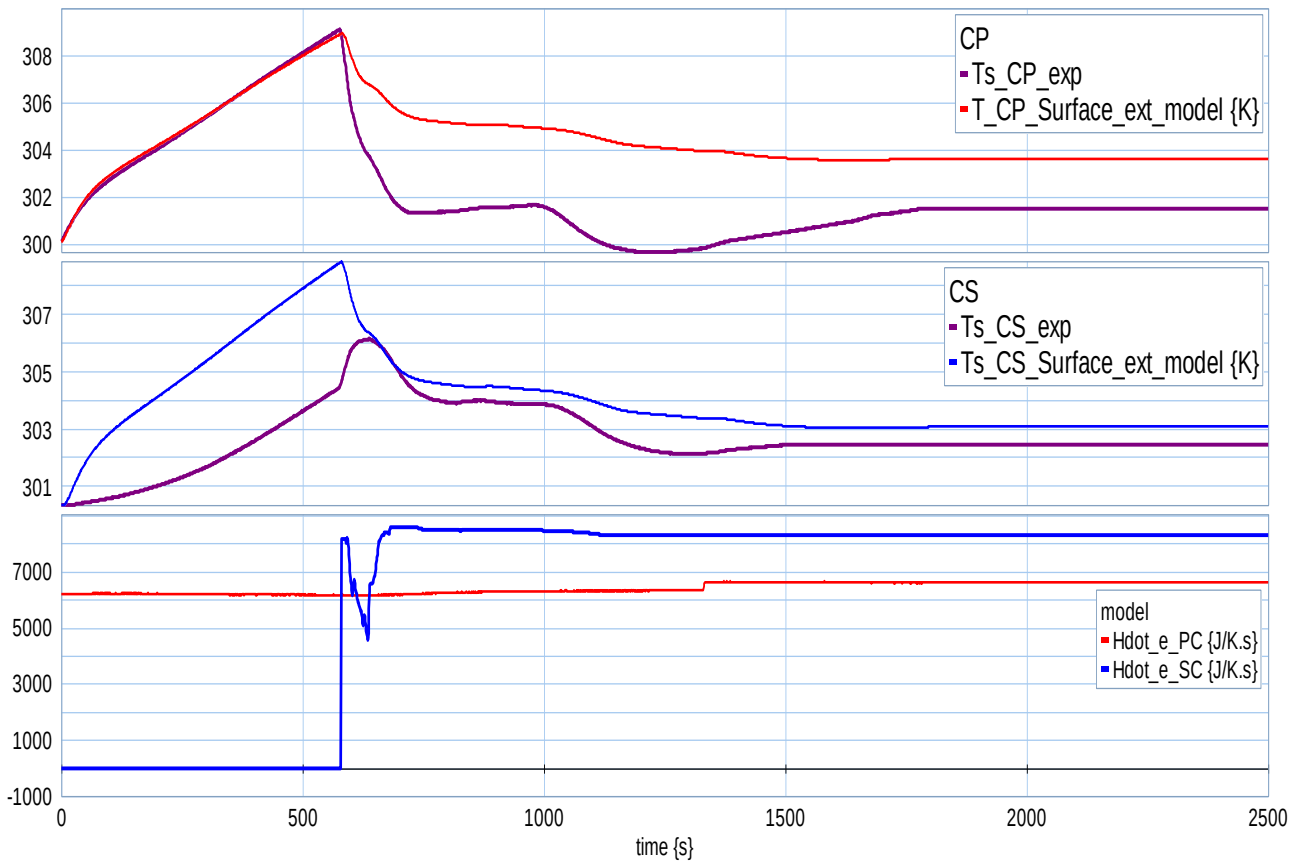

***Fig.9****: Temperatures profiles in both PC and SC with* $h_{PC} = h_{SC} = 6530$ w/m²°C.

## IV. Identification and optimization of the convective conductance.

The accurate determination of the heat transfer coefficient "h" inside the condenser is very difficult because of the complexity of the corrugated geometry of the exchanger. However the dynamic BG model proposed in this study is equipped with a semi empirical model for its estimation. The objective of the current section is to optimize the thermal conductance in both compartments of the condenser PC and SC. Starting with the nominal value "$h_0$" calculated by the proposed semi empirical model (eq.9) for each compartment, then an optimization is performed on "$h_0$" to minimize the difference in temperatures between the model and experience.

The optimization used in this study is based on Broydon Fletcher Goldfarb Shanno method, which is already integrated into the 20sim simulation software**.** This method uses both the gradient of a function and the second order gradient to determine the search direction. The search direction is kept for each new step until a minimum has been found. Then a new search direction is determined and the process goes on.

### a) First test with a constant conductance value.

The brazed-plate heat exchanger used in our study is delivered with data sheet where the manufacturer mentions a value of **h = 6530** w/m²°C. The idea here is to launch the simulation of the first test case (**4.a**) with this imposed value and compare the profile of simulating temperatures with experimental measurements (**Fig.9**).

The dynamic model simulated with this value of "h" shows a discrepancy of **2°C** with the measured temperature in steady state. This means that the imposed value of conductance does not correspond to the right value.

### a) Improvement of the convective conductance.

The scientific literature describes very few models of the convective conductance specific to brazed-plate heat exchanger. However, Alfa Laval is a condenser manufacturer which discloses a model adapted to our current application and whose formula is:

$$h = K \times \lambda \times P_r^{1/3} \times (\frac{\rho \times \Delta P}{\mu^2})0{,}3274 \qquad (9)$$

with:

| | |
|---|---|
| $h$ : | Heat transfer coefficient (in w/m²°K *)* |
| $K$ : | Optimization parameter, initially equal to 234. |
| $\Pr$ : | Prandtl number (by definition $Cp * \mu / \lambda$ ). |
| $\Delta P$ : | Drop pressure (in kPa) |
| $\mu$ : | Viscosity (in cP). (1Ns/m² = $10^3$cP). |

***Test 1:***

Simulation of the case detailed in paragraph (**III.a**), is run again, however, with the heat transfer coefficient *"h"* evaluated using AlfaLaval equation. The latter is then optimized according to Broydon method. Optimization of the two coefficients "h" (PC and SC) converges to the following values:

$$h_{opt_PC} = 1062 \text{ w/m}^2{}^\circ\text{K}, \quad h_{opt_SC} = 4152 \text{ w/m}^2{}^\circ\text{K}.$$

**Fig.10** shows a better concordance between numerical resolution (red and blue curves) and experience (purple curves). The difference being recorded is reduced from **2°C** to **1.2 °C** in primary circuit, and from **0.9°C** to **0.2°C** in secondary circuit, which is inside the precision domain of the temperature sensors. The result is slightly better for the SC because the fluid is pure water, even though the fluid in the hot circuit (PC) is not really pure water but contains a certain quality of oil for the pump lubrication.

***Test 2:***

*Initial conditions:*

- Cold mass flow rate = 4 cm3/s.
- Hot mass flow rate = 0 cm3/s.

At the hot circuit, we notice that at the steady state regime, the temperature difference between the curve from the model and the experimental is around 1 degree, which may correspond to the uncertainty of the thermocouples. Whereas at cold circuit, we clearly see that the two curves are superimposed (***fig.12***).

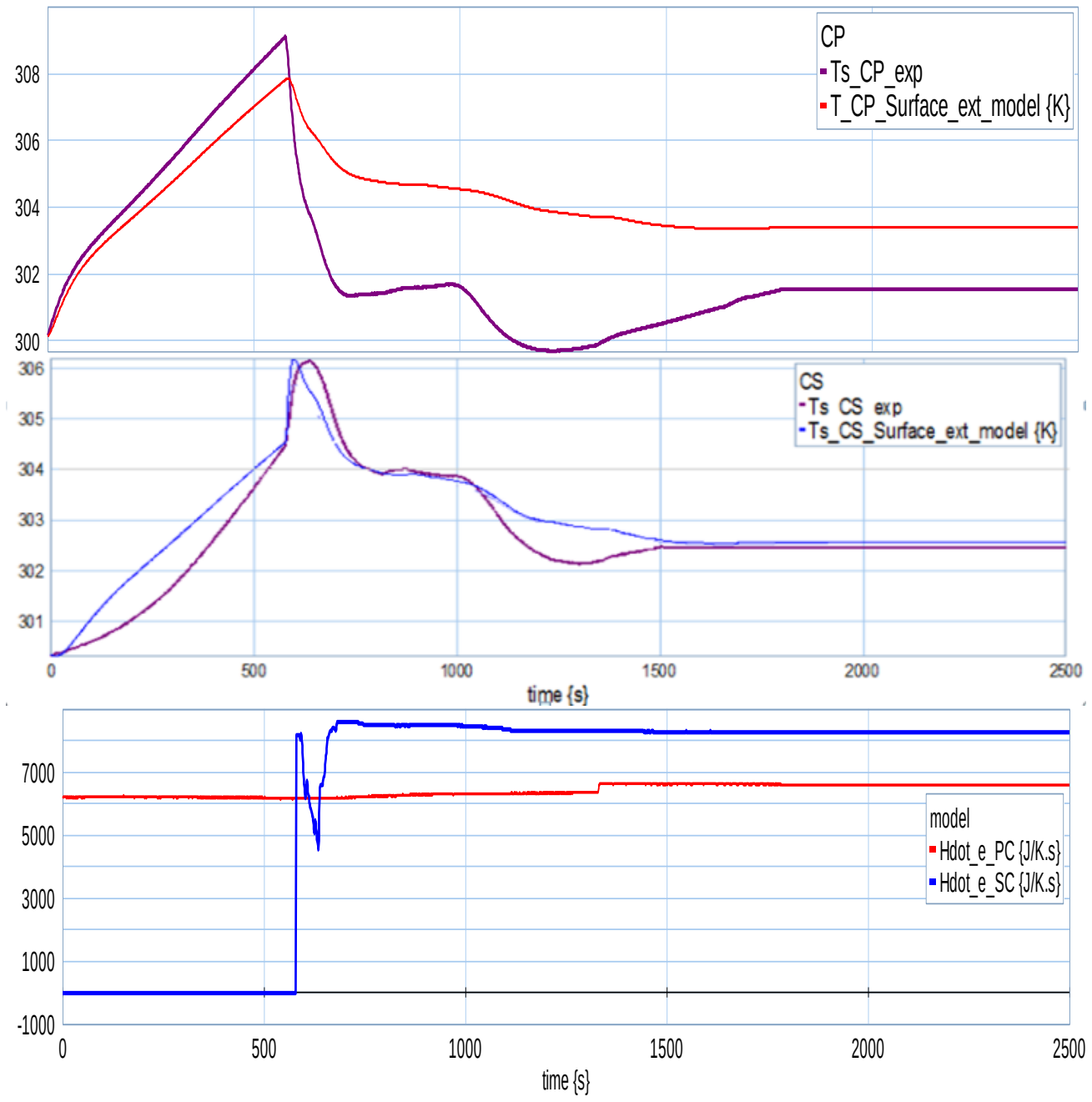

***Fig.10***: *Temperatures profiles in both PC and SC with Optimized Alfa Laval "h".*

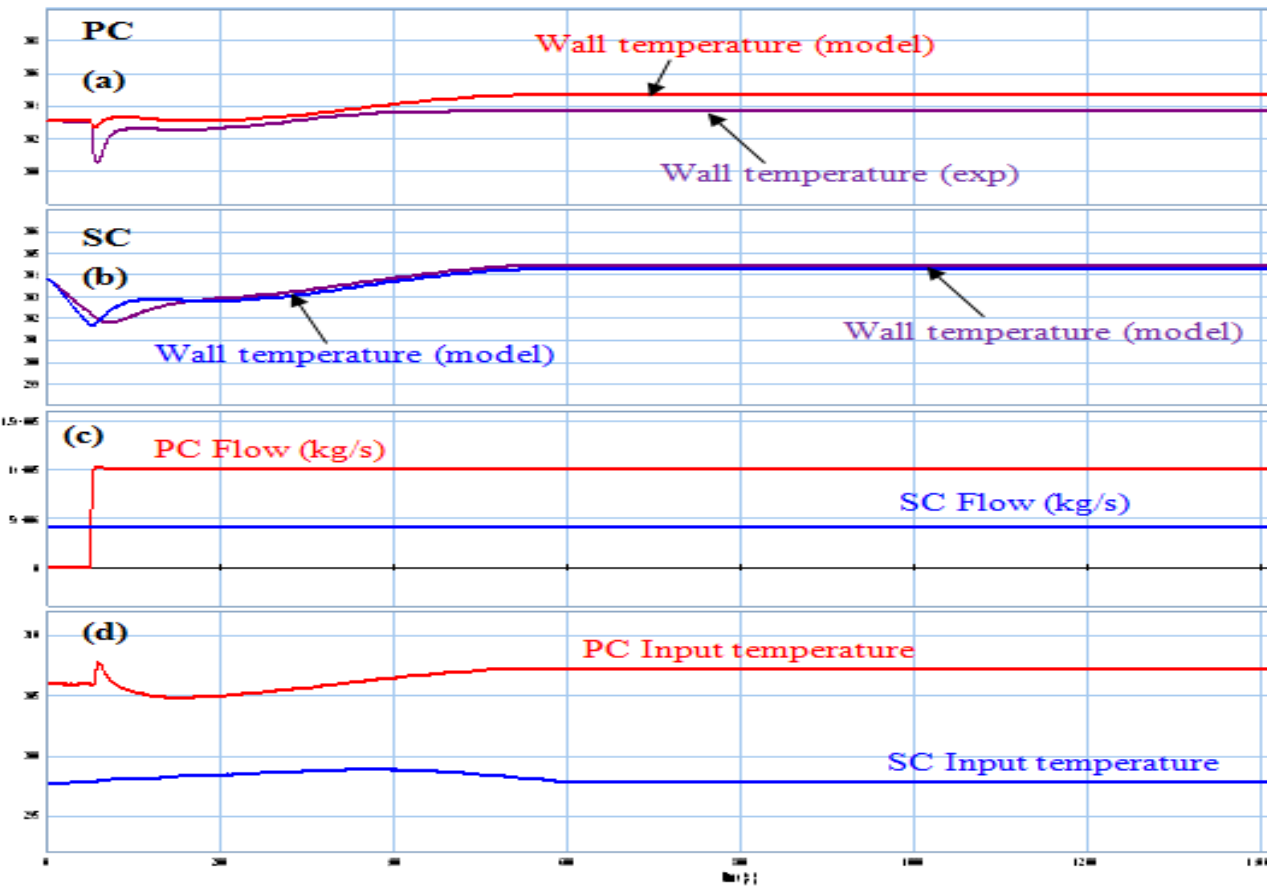

***Fig.12:*** *Time evolution of fluid temperature in PC (a), in SC (b), flow evolution in both PC and SC (c) and input temperature in both PC and SC (d).*

## V- Conclusion

In this paper we propose a Brazed-Plate heat exchanger model based on BG approach, taking into consideration the hydro-thermal exchanges that occur in the system. Due to simplified assumptions, the difference between theoretical and experimental results can be attributed to the uncertainty of the sensors. The model can be considered as reliable enough to represent the heat transfer in the heat exchanger in monophasic behavior. The model is parametrized in terms of device geometry, type of fluid, input variables (hot and cold sources and temperatures). The future research work is to model the two-phase regime and validate it through experiments.

## Model variables.

| | | |
|---|---|---|
| $C_p$ | Specific heat | J/kg/K |
| $C$ | Capacity | J/K |
| $e$ | Channels thickness | m |
| $H$ | Enthalpy | J |
| $\dot{H}$ | Heat flow | J/s |
| $H_e$ | Inlet enthalpy | J/kg |
| $Hs$ | Outlet enthalpy | J/kg |
| $L$ | Channels length | m |
| $m$ | Mass | kg. |
| $P$ | Pressure | Pa |
| $q_m$ | Mass flow | kg/s |
| $\dot{Q}_{ap}$ | Heat flow (Hot source) | J/s |
| $R$ | Resistance. | |
| $T$ | Temperature | K |
| $V$ | Volume | m3 |
| ρ | Density of the fluid | kg/m3 |
| $v$ | Velocity | m/s |
| μ | Viscosity | Pa.s |
| $m$ | Mass. | kg |
| 0 | Initial state. | |
| liq | Liquid. | |
| cond | Condenser. | |
| $\lambda$ | Thermal conductivity of water | |
| MSe: | Modulate source effort. | |

## Acknowledgment

This paper describes results from research supported by the FUI 14 Program in the context of project ThermoFluid RT labeled by the competitivity pole ASTech. The authors gratefully acknowledge Mr R. Albach from Atmostat company for his important contribution to the experimentation phase, and the members of the consortium for the multiple fruitful scientific discussions about heat exchanges.

## Reference

[1] M. Kebdani, G. Dauphin-Tanguy, A. Dazin, P. Dupont *Bond Graph Model of a mechanically Pumped Biphasic Loop (MPBL)*, 23rd Mediterranean Conference on Control and Automation, Meliá Costa del Sol, Torremolinos, Spain, 16-19 Juin, 2015.

[2]http://www.swep.net/fr/products_solutions/solutions/Pages/condensers.aspx

[3] Investigation des transferts thermiques locaux dans un échangeur à plaques par thermographie infrarouge Kifah SARRAF1, Stéphane LAUNAY1, Lounès TADRIST1 Aix-Marseille Université, Laboratoire IUSTI, UMR CNRS 7343, Marseille cedex 13, France.

[4] Cremaschi, L., A. Barve, and X. Wu. 2012. Effect of Condensation Temperature and Water Quality on Fouling of Brazed-Plate Heat Exchangers. *ASHRAE Transactions.* 118(1):1086-1100.

[5] Focke, W.W., J. Zachariades, and I. Olivier. 1985. Effect of the corrugation inclination angle on the thermohydraulic performance of plate heat exchangers. *International Journal of Heat and Mass Transfer* 28:1469–1479.

[6] Vlasogiannis, P., Karagiannis, G., Argyropoulos, P. & Bontozoglou, V., "Air-water two-phase flow and heat transfer in a plate heat exchanger", International Journal of Multiphase Flow, 28(5), pp. 757-772, 2002.

[7] Volker, G. & Kabelac, S., "Experimental investigations and modelling of condensation in plate heat exchangers", Washington, s.n., 2010.

[8] S. Freund and S. Kabelac, "Investigation of local heat transfer coefficients in plate heat exchangers", International Journal of Heat and Mass Transfer, vol. 53, p. 3764–3781, 2010.

[9] X. Rong, M. Kawaji and J.G. Burgers, Two-phase header flow distribution in a stacked plate heat exchanger, *Proceedings ASME/JSME FED-Gas Liquid Flows* 225 (1995), pp. 115–122.

[10] Cooper, A., Usher, J.D., 1983. Plate heat exchangers. In: Schlunder, E.U. (Ed.), Heat Exchanger Design Handbook, vol. 3. Hemisphere, Washington.

[11] Raju, K.S.N., Bansal, J.C., 1983. Design of plate heat exchangers. In: Kakac, S., et al. (Eds.), Low Reynolds Number Flow Heat Exchangers. Hemisphere, Washington, pp. 913–932.

[12] Focke, W.W., Zachariades, J., Olivier, I., 1985. The effect of the corrugation inclination angle on the thermohydraulic performance of plate heat exchangers. Int. J. Heat Mass Transf. 28, 1469–1497.

[13] Focke, W.W., Knibbe, P.G., 1986. Flow visualization in parallel-plate ducts with corrugated walls. J. Fluid Mech. 165, 73–77.

[14] Bansal, B., Muller-Steinhagen, H., 1993. Crystallization fouling in plate heat exchangers. ASME J. Heat Transfer 115, 584–591.

[15] Ould Bouamama, B., Thoma, J. U., Cassar, J.P. (1997). *Bond Graph modelisation of steam condenser*. 0-7803-4053-1/97/$10.00. Article IEEE. Automatic Control, Computer Engineering and Signal Laboratory